\documentclass[manuscript,11pt]{aastex}
\usepackage{epsfig}
\usepackage{soul}
\usepackage{color}

\usepackage{physymb}
\usepackage{mathrsfs}

\def\fexxv{Fe\,{\sc xxv}}
\def\fexxvi{Fe\,{\sc xxvi}}

\def\mathv{\textbf{\em v}}
\def\mathB{\textbf{\em B}}

\def\mathJ{\textbf{\em J}}
\def\mathn{\textbf{\em n}}

\def\cm{\ifmmode {\rm cm}^{-1} \else cm$^{-1}$ \fi}
\def\s{\ifmmode {\rm s}^{-1} \else s$^{-1}$ \fi}
\def\cc{\ifmmode {\rm cm}^{-3} \else cm$^{-3}$ \fi}
\def\cs{\ifmmode {\rm cm}^{-2} \else cm$^{-2}$ \fi}
\def\g{\ifmmode \gamma \else $\gamma$\fi}
\def\G{\ifmmode \Gamma \else $\Gamma$\fi}
\def\Gs{\ifmmode \Gamma~ \else $\Gamma~$\fi}

\def\gc{\ifmmode \gamma_{\rm c} \else $\gamma_{\rm c}$ \fi}

\def\gsim{\mathrel{\raise.5ex\hbox{$>$}\mkern-14mu
             \lower0.6ex\hbox{$\sim$}}}
\def\lsim{\mathrel{\raise.3ex\hbox{$<$}\mkern-14mu
             \lower0.6ex\hbox{$\sim$}}}
\def\simless{\mathbin{\lower 3pt\hbox
     {$\rlap{\raise 5pt\hbox{$\char'074$}}\mathchar"7218$}}}   
\def\simmore{\mathbin{\lower 3pt\hbox
     {$\rlap{\raise 5pt\hbox{$\char'076$}}\mathchar"7218$}}}   
\def\Msun{M_\odot}                                
\def\4u{4U 1728--34}
\def\deg{^\circ}

\def\obj{PDS~456}

\lefthead{et al.} \righthead{\4u}

\shorttitle{Coronal Broadening of UFOs}

\shortauthors{Fukumura et al.}

\begin{document}

\title{Constraining X-ray Coronal Size with Transverse Motion of AGN Ultra-Fast Outflows}

\date{\today}

\author{\textsc{Keigo Fukumura}\altaffilmark{1},
\textsc{and} \textsc{Francesco Tombesi}\altaffilmark{2,3,4,5}}
\altaffiltext{1}{Department of Physics and Astronomy, James Madison University,
Harrisonburg, VA 22807; fukumukx@jmu.edu}
\altaffiltext{2}{Astrophysics Science Division, NASA/Goddard Space Flight Center,
Greenbelt, MD 20771}
%
%
\altaffiltext{3}{Department of Astronomy, University of Maryland, College
Park, MD20742}
\altaffiltext{4}{Department of Physics, University of Rome ``Tor
Vergata", Via della Ricerca Scientifica 1, I-00133 Rome, Italy}
\altaffiltext{5}{INAF Astronomical Observatory of Rome, Via Frascati 33, 00078 Monteporzio Catone (Rome), Italy}

\begin{abstract}
\baselineskip=15pt

One of the canonical physical properties of ultra-fast outflows (UFOs) seen in a diverse population of active galactic nuclei (AGNs) is its seemingly very broad width (i.e. $\Delta v \gsim 10,000$ km~s$^{-1}$) , a feature often required for X-ray spectral modeling. While unclear to date, this condition is occasionally interpreted and justified  
as 
internal turbulence within the UFOs for simplicity. In this work, we exploit a transverse motion of a three-dimensional  accretion-disk wind, an essential feature of non-radial outflow morphology unique to magnetohydrodynamic (MHD) outflows. We argue that at least part of the observed line width of UFOs may  reflect the  degree of transverse velocity gradient  due to Doppler broadening around a putative compact X-ray corona in the proximity of a black hole. In this scenario, line broadening is sensitive to the geometrical size of the corona, $R_c$. 
We calculate the broadening factor as a function of coronal radius $R_c$ and velocity smearing factor $f_{\rm sm}$ at a given plasma position. 
We demonstrate, as a case study of the quasar, PDS~456, that the spectral analysis favors a compact coronal size of $R_c /R_g \lesssim 10$ where $R_g$ is gravitational radius.
Such a compact corona is long speculated from both X-ray reverberation study and the lamppost model for disk emission also consistent with microlensing results. 
Combination of such a transverse broadening around a small corona can be a direct probe of a substantial rotational motion perhaps  posing a serious challenge to radiation-driven wind viewpoint.

\end{abstract}



\baselineskip=15pt

\section{Introduction}

Ultra-fast outflows (UFOs) are an intriguing sub-class of ionized outflows in the form of resonant absorption lines seen primarily by high-throughput CCD observations  in about 40\% of active galactic nuclei (AGNs)  \citep[][]{Tombesi10, Tombesi14,Tombesi15,Gofford13} including very bright (lensed) quasars \citep[QSOs;][]{Chartas03, Reeves03, Reeves18, Lanzuisi12, Pounds03, Chartas09, Vignali15, Dadina18}. They are detected also in non-AGN sources such as ultra-luminous X-ray sources \citep[][]{Walton16} and potentially tidal disruption events \citep[e.g.][]{Kara18}. In general, canonical X-ray UFOs\footnote[1]{Note also that there have been a handful reports implying non-canonical UFOs with low ionization state ($\log \xi \sim 0-2$) and/or low column density ($N_H \sim 10^{20} - 10^{22}$ cm$^{-2}$), but this is beyond the scope of the present work \citep[e.g.][]{Gupta13,Longinotti15,Serafinelli19}.} are known to exhibit extreme physical conditions; namely, a massive column density ($N_H \gsim 10^{23}$ cm$^{-2}$), highly ionized state (of ionization parameter\footnote[2]{This is defined as $\xi \equiv L_{\rm ion} / (n r^2)$ where $L_{\rm ion}$ is ionizing (X-ray) luminosity, $n$ is plasma number density at distance $r$ from the BH.}  of $\log \xi \sim 4-6$) outflowing at near-relativistic speed ($v/c \sim 0.1-0.7$). Because of its energetically powerful nature, UFOs can be good candidates to effectively deliver sufficient amount of energy and momentum out to host galaxies at $\sim$ kpc-scale to  quench star formation activity \citep[e.g.][]{Hopkins10,Tombesi15}.

Quasar outflows generally have been thought to be driven by radiation
pressure by their intense O/UV flux, in a manner similar to that observed in massive stars
\citep[][]{KingPounds15,Hagino17,Nomura17}. On the other hand, most UFOs are 
highly ionized such that little UV or  soft X-ray opacity may be left in the wind, making this
process very inefficient \citep[e.g.][]{Higginbottom14}. In addition, the production region of line-driven UFOs seems to be spatially limited to a very narrow latitudinal region (e.g. $70\deg \lesssim \theta \lesssim 80\deg$) according to  hydrodynamic simulations \citep[e.g.][]{Nomura17}, which casts a doubt in viability with observations. 
Also, a number of UFOs have been found in sub-Eddington accretors \citep[e.g.][]{Marinucci18}. To overcome these potential issues, another plausible mechanism -- magnetically-driven disk-winds -- has been proposed  to provide
an alternative means to efficiently accelerate the highly ionized absorbers as observed (e.g.
\citealt{BP82}, hereafter, BP82; \citealt{CL94}, hereafter, CL94; \citealt{KK94}; \citealt{Ferreira97}; 
\citealt{F10}, hereafter F10; \citealt{K12}; \citealt{K19}; \citealt{Kraemer18}; \citealt{F18}, hereafter F18) in the context of both AGNs and X-ray binaries. MHD wind in general is thus immune to these issues.


Another puzzle  associated with the canonical UFOs is its remarkably broad line width (e.g. $\sigma_{\rm tur} \sim 10,000-20,000$ km~s$^{-1}$) often required for photoionization modeling, for example, with {\tt xstar} that employs an arbitrary choice of internal turbulence parameter (aka. {\tt vturb}; e.g. \citealt{Nardini15}, hereafter N15; \citealt{Tombesi15}; \citealt{Reeves18}). 
Although broadening is phenomenologically interpreted as turbulence presumably intrinsic to the outflowing plasma, such a large velocity dispersion is not easily conceivable from first principle. More specifically, wind morphology would be essentially one-dimensional (i.e. radial) in the context of radiation-driven scenario, which would leave little room for causing velocity gradient, thus 
necessarily attributed to something like turbulence. On the other hand, it is more realistic to consider that the broadening might represent the actual wind kinematic field. 
In F18, for example, line broadening is naturally determined by radial velocity shear ($\Delta v_r$) of the wind along a line of sight (LoS) without using {\tt vturb}. 
Alternatively, magnetically launched winds are known to possess substantial azimuthal velocity component, $v_\phi$, originating from the accretion disk surface (see BP82; CL94; F10). 
This unique MHD-wind feature has prompted us to exploit the role of the transverse motion in characterizing UFO spectrum in this paper.

   
\obj\ is a well-studied nearby ($z=0.184$), radio-quiet QSO hosting a black hole (BH) of mass $M \sim 10^9 \Msun$ \citep[e.g.][]{Reeves09}, being the most luminous AGN in the local universe with a bolometric luminosity of $L_{\rm bol} \sim 10^{47}$ erg~s$^{-1}$. It is among the best known QSOs to exhibit archetypial UFO signatures  in the Fe K band from the past X-ray observations with {\it Suzaku} \citep[e.g.][hereafter M17]{Reeves09,Gofford14,Matzeu17}, {\it XMM-Newton}/EPIC and {\it NuSTAR} (\citealt[e.g.][]{Behar10}; N15) since its first discovery \citep[e.g.][]{Reeves03}. 
Extensive spectral analyses so far have systematically revealed the existence of   highly ionized UFOs of $v/c \sim 0.2-0.3$ identified as the 1s-2p resonance transitions of \fexxvi\ with  $N_H \lesssim 10^{24}$ cm$^{-2}$ and high ionization parameter of $\log \xi \sim 4-6$ where
%
the detected UFOs may well be located in a close proximity to the BH within $\sim 100 R_g$ \citep[e.g.][]{Reeves09}. 
Recently, even a faster UFO component has been reported in PDS~456 \citep[e.g.][]{Reeves18,Boissay19}.


Motivated by these observations and ideas, we attribute in this paper the line width to a significant transverse motion naturally expected in the MHD-wind framework that we have developed earlier. 
We show how line broadening is dependent on the size of the corona in this model.
In \S 2, we overview a physical setup of the problem  by introducing the magnetically-launched disk-wind model. In \S 3, we present our calculations and the bestfit result of \obj\ in an attempt to constrain the size of the ionizing X-ray corona. Finally in \S 4, we conclude with a summary and discussion.

\begin{figure}[t]
\begin{center}$
\begin{array}{cc}
\includegraphics[trim=0in 0in 0in
0in,keepaspectratio=false,width=4.0in,angle=-0,clip=false]{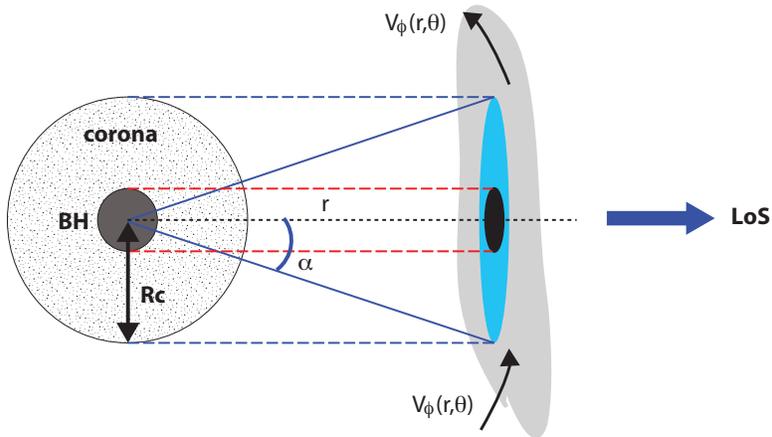}
\end{array}$
\end{center}
\caption{A face-on schematic picture illustrating a transverse wind motion  of the plasma (blue region) around a BH corona  intersecting a line of sight of a distant observer.}
\label{fig:f1}
\end{figure}

\section{Model Description}

\subsection{Magnetized Winds}

Considering pure magnetic $\mathJ \times \mathB$ force, both magnetic pressure primarily from toroidal field $\propto \grad \mathB^2$ as well as tension due to $(\mathB \cdot \grad) \mathB$, can successfully drive an UFO of massive column with the observed relativistic speed  being photoionized to have high $\xi$ values (see F10 and F18 for more detailed model description; reference therein).  
%
We employ the self-similar prescription in the radial direction assuming the escaping velocity profile of $v(r,\theta) \propto r^{-1/2} f(\theta)$ where $f(\theta)$ denotes the angular dependence to be calculated and  the poloidal field structure is determined by numerically solving the MHD Grad-Shafranov equation as is originally formulated in CL94 and applied in F10. 


\subsection{Line Broadening due to Transverse Motion}

The plasma is initially differentially rotating in the disk around a BH in a proximity to a putative X-ray corona responsible for photoionizing the surrounding materials. That is, the wind is initially transversing in front of the central corona with respect to a distant observer under axisymmetry, $v_\phi(r,\theta)$. Then, the change in $v_\phi$ is further dependent on the physical size of the corona near the BH. Consequently, line broadening due to transverse motion in this picture is determined by the coronal size. To quantify the degree of broadening factor,  we construct a simplistic toy model as illustrated in {\bf Figure~\ref{fig:f1}}, in an attempt to further calculate the effect on UFO line spectrum. The central corona is assumed to be a spherical region of radius $R_c$ surrounding the BH. While winds 
are being photoionized isotropically by the corona, classical Doppler shift in velocity depends primarily on (1) the cross sectional size of the corona, $R_c$, that intersects with a LoS of a distant observer and (2) the absorber's distance, $r$. 

If the line width is primarily attributed to the transverse motion of the wind under illumination  directly in front of the corona (blue region), then the classical azimuthal Doppler shift $\Delta v$ is expressed by
\begin{eqnarray}
\frac{\Delta v(r,\theta,\phi)}{v_c} \equiv \frac{\mathv_w \cdot \mathn}{c}  =  \frac{v_\phi(r,\theta;r_o)}{c} \sin \theta_{\rm obs} \sin \phi  \  ,\label{eq:shift1}
\end{eqnarray}
where $\mathn$ is a unit vector along a line of sight (LoS) in the spherical coordinates $(r,\theta,\phi)$, $r_o$ is a characteristic innermost launching radius and $v_c$ denotes a centroid (unmodulated) LoS velocity. 
Note that the shift due to the poloidal velocity component, $v_p \hat{n_p}$, produces no modulation and thus is ignored here. 

Considering that the observed line broadening of UFOs is reflecting the background irradiating coronal size (i.e. radius) as depicted in {\bf Figure~\ref{fig:f1}}, the solid angle subtended from the center of the corona by the wind plasma at radius $r$ (corresponding to the blue region) yields a relation between the radius $R_c$ and the toroidal angle $\alpha$ such that 
\begin{eqnarray}
-\frac{R_c}{r} \le \alpha \le \frac{R_c}{r} \  . \label{eq:angle}
\end{eqnarray}
Therefore, by substituting $\alpha$ into $\phi$,  we note that $\Delta v_i = \Delta v(r,\theta,\phi_i; r_o,R_c)$ where $-\alpha \lesssim \phi_i \lesssim \alpha$ is the $i^{\rm th}$ azimuthal position of the discretized wind. We then calculate its root-mean-square (rms) value 
\begin{eqnarray}
\Delta v_{\rm rms} = \sqrt{\sum_i^N (\Delta v_i)^2 /N_\phi } \  . \label{eq:Vrms}
\end{eqnarray}
for a characteristic transverse modulation that dictates the  line broadening where $N_\phi$ is the discrete number of the azimuthal plasma position within the angle $\alpha$. To further  accommodate an external smearing effect in transverse motion, we parameterize the effective broadening by introducing a smearing factor $f_{\rm sm}$ as $\Delta v_{\rm eff} \equiv f_{\rm sm} \Delta v_{\rm rms}$ where $0.1 \lesssim f_{\rm sm} \lesssim 1$ is assumed.

\subsection{Spectral Modeling of Fe K UFOs in PDS~456 }

We apply the model described above to 
the simultaneous observations of PDS~456 with {\it XMM-Newton} and {\it NuSTAR} in 2013/2014 (see, e.g., N15 and F18). Our current approach directly follows the methodology adopted in F18. For photoionization calculations with {\tt xstar} \citep[][]{KallmanBausas01}, we use the same library of MHD wind solutions as used in F18 assuming a double broken power-law ionizing spectrum. 
The local ionic column $N_{\rm ion}$ is computed with {\tt xstar}  under thermal equilibrium, while the photo-absorption cross section $\sigma_{\rm abs}$ is calculated using
the usual Voigt profile as a function of photon frequency $\nu$ and  the line broadening factor  $\Delta \nu_\ell \approx  (\Delta v_{\rm eff}/c) \nu_c$ relative to the centroid (rest-frame) frequency $\nu_c$. In F18, the line broadening was attributed to the LoS (radial) shear velocity of the wind $\Delta v_r$. In this paper, however, we consider that the transverse (toroidal) motion of the wind primarily contributes to line broadening, as described  in \S 2.1. One can then calculate the line optical depth  $\tau_\ell = \sigma_{\rm abs} N_{\rm ion}$ to simulate the spectrum. To reduce the degree of freedom in the problem, we adopt some of the model parameters obtained from F18; i.e. LoS angle of $\theta=50\deg$ with the X-ray luminosity of $L_{\rm ion}
\sim 5 \times 10^{44}$ erg/s in this specific epoch (see also \citealt{Gofford14}; N15; M17) for the wind density profile of $n \propto r^{-1.2}$, although the last assumption is less relevant for the current UFO modeling. Thus, our  model in this work has three free parameters that essentially governs the predicted spectrum; density $n_{10}$ in the units of $10^{10}$ [cm$^{-3}$] of the wind at the innermost launching radius, radius of the corona $R_c$ and smearing factor $f_{\rm sm}$.

\begin{figure}[h]
\begin{center}$
\begin{array}{cc}
\includegraphics[trim=0in 0in 0in
0in,keepaspectratio=false,width=2.9in,angle=-0,clip=false]{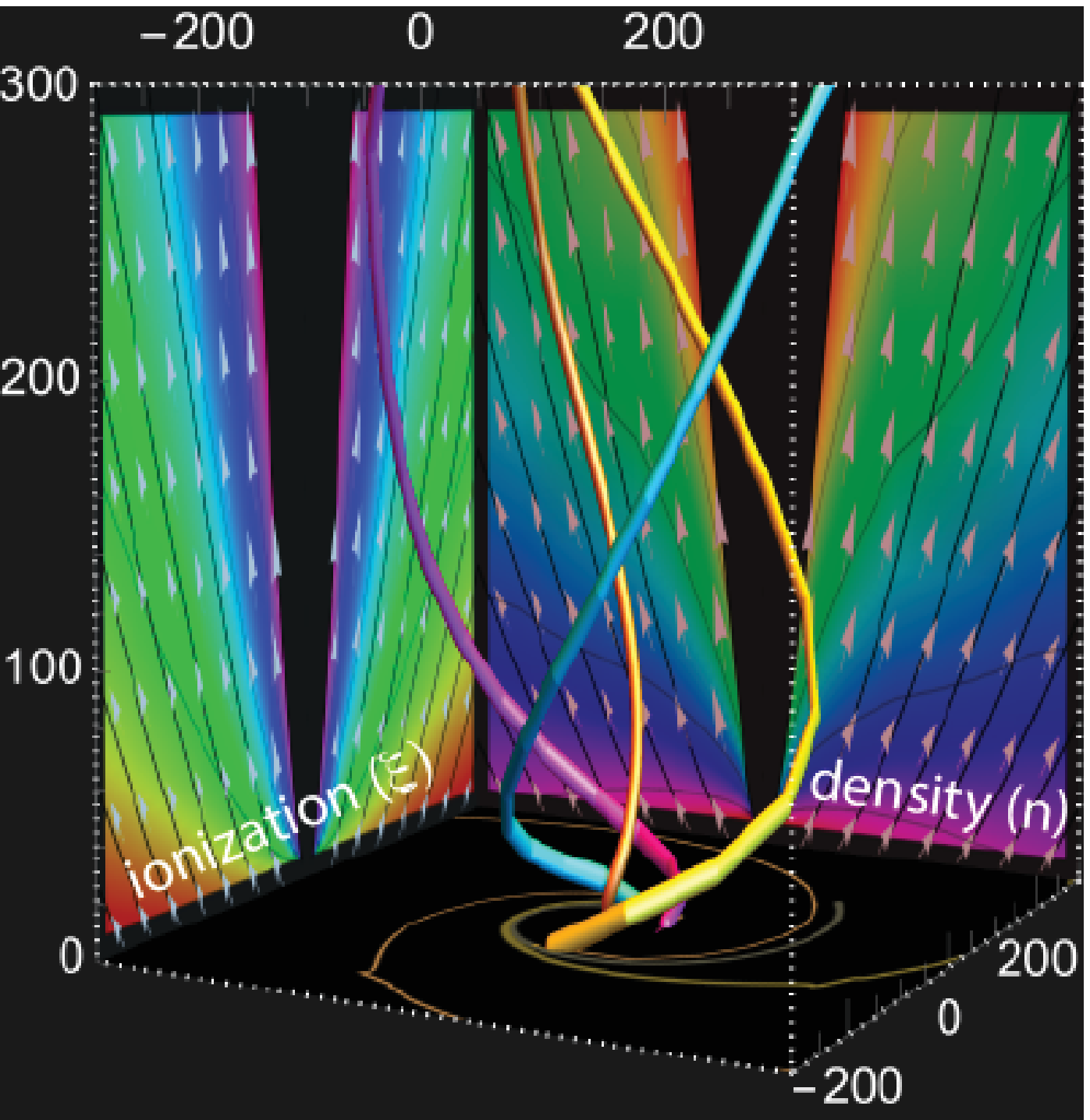}
\includegraphics[trim=0in 0in 0in
0in,keepaspectratio=false,width=3.6in,angle=-0,clip=false]{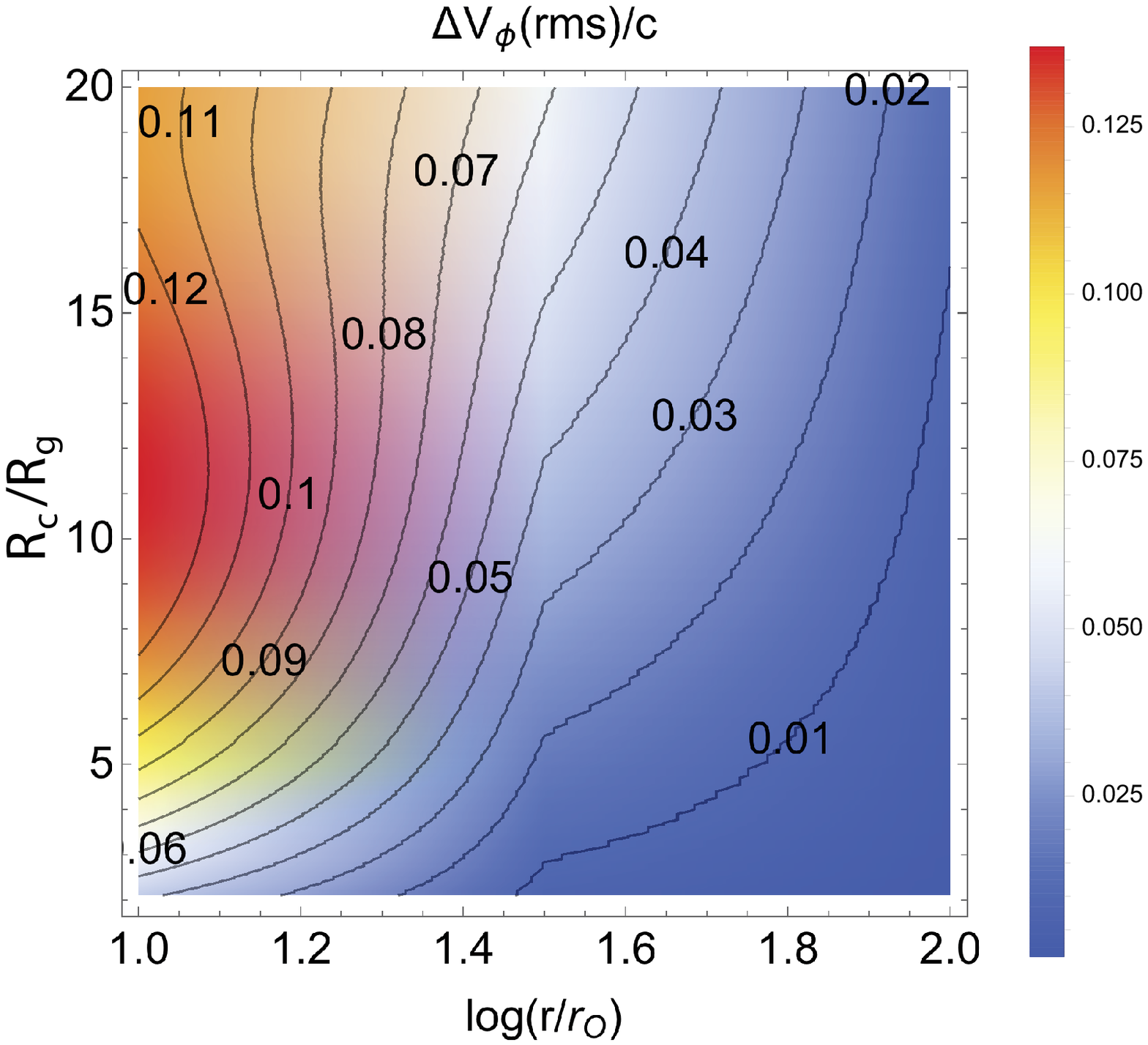}
%
\end{array}$
\end{center}
\caption{(a) A three-dimensional rendering of the streamlines of a fiducial MHD-driven disk-winds considered for the spectral modeling. Normalized wind density $n(r,\theta)$ and ionization parameter $\xi(r,\theta)$ are shown where magenta/dark blue denotes highest value and yellow/red shows lowest values (roughly ranging over 6 orders of magnitudes). Transverse motion especially at small radii is one of the generic features of the MHD winds. (b) Transverse Doppler broadening factor $\Delta v_{\rm rms}/c$ 
calculated from a fiducial wind solution (from F18) as a function of LoS radial position $r$ and radial coronal extent $R_c$ for $\theta_{\rm obs}=50\deg$.   } \label{fig:wind}
\end{figure}

\section{Results}

\subsection{Dependences of Coronal Size}

By calculating the broadening effect due to the toroidal motion  as a function of $R_c$, we first calculate the distribution of $\Delta v_{\rm eff}$ as a function of $R_c$ and the wind position $r$ for a given LoS (set to be $50\deg$ here as treated in F18).  A global morphology of the calculated MHD disk-wind is presented in {\bf Figure~\ref{fig:wind}a} showing  its normalized wind density $n(r,\theta)$ and ionization parameter $\xi(r,\theta)$ roughly ranging over 6 orders of magnitudes. Among the generic features of the MHD wind is dominant toroidal rotation near the launching radius, which is then quickly transformed into poloidal motion as the wind becomes accelerated. It is calculated that the wind possesses a significant transverse motion especially at small radii $r$, which in our current work is responsible for line broadening. Note that the wind in this domain (i.e. $r/R_g \lesssim 100$) is highly ionized at high velocity that is important for UFOs. 

Being emphasized on the toroidal motion, in {\bf Figure~\ref{fig:wind}b} we also calculate the distribution of the line energy shift due to $\Delta v_{\rm rms}$ determined by equation~(\ref{eq:Vrms}) as a function of $R_c$ and wind radius $r$. For $\theta=50\deg$ as considered here and in F18, the maximum broadening  can be up to $\Delta v_{\rm rms}/c \lesssim 0.1$ for $10 \lesssim r/R_g \lesssim 100$ if $f_{\rm sm}=1$, which is sufficient to account for the observed line width seen in PDS~456. As $R_c$ increases, a larger fraction of the wind with a wider opening angle comes into the LoS of a distant observer.  Hence, $\Delta v_{\rm rms}$ increases with $R_c$ gradually at a given radius $r$. On the other hand, the toroidal velocity declines with increasing radius as $v_{\phi} \propto r^{-1/2}$ for a given coronal size. Given the expected distribution of the broadening factor as shown here, one can calculate line spectrum as discussed  in \S 3.2.

\begin{figure}[h]
\begin{center}$
\begin{array}{cc}
\includegraphics[trim=0in 0in 0in
0in,keepaspectratio=false,width=3.45in,angle=-0,clip=false]{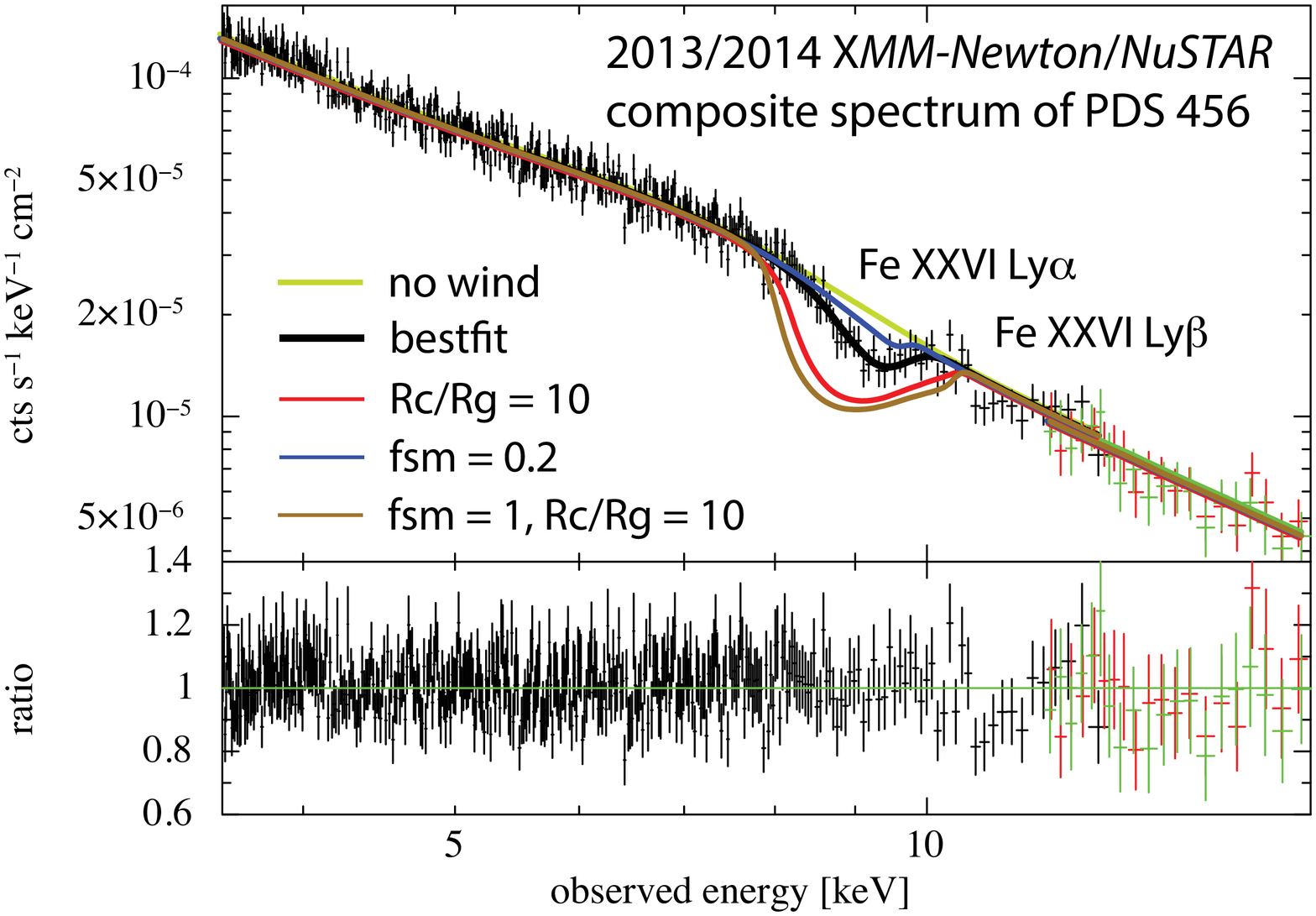}
\includegraphics[trim=0in 0in 0in
0in,keepaspectratio=false,width=3.2in,angle=-0,clip=false]{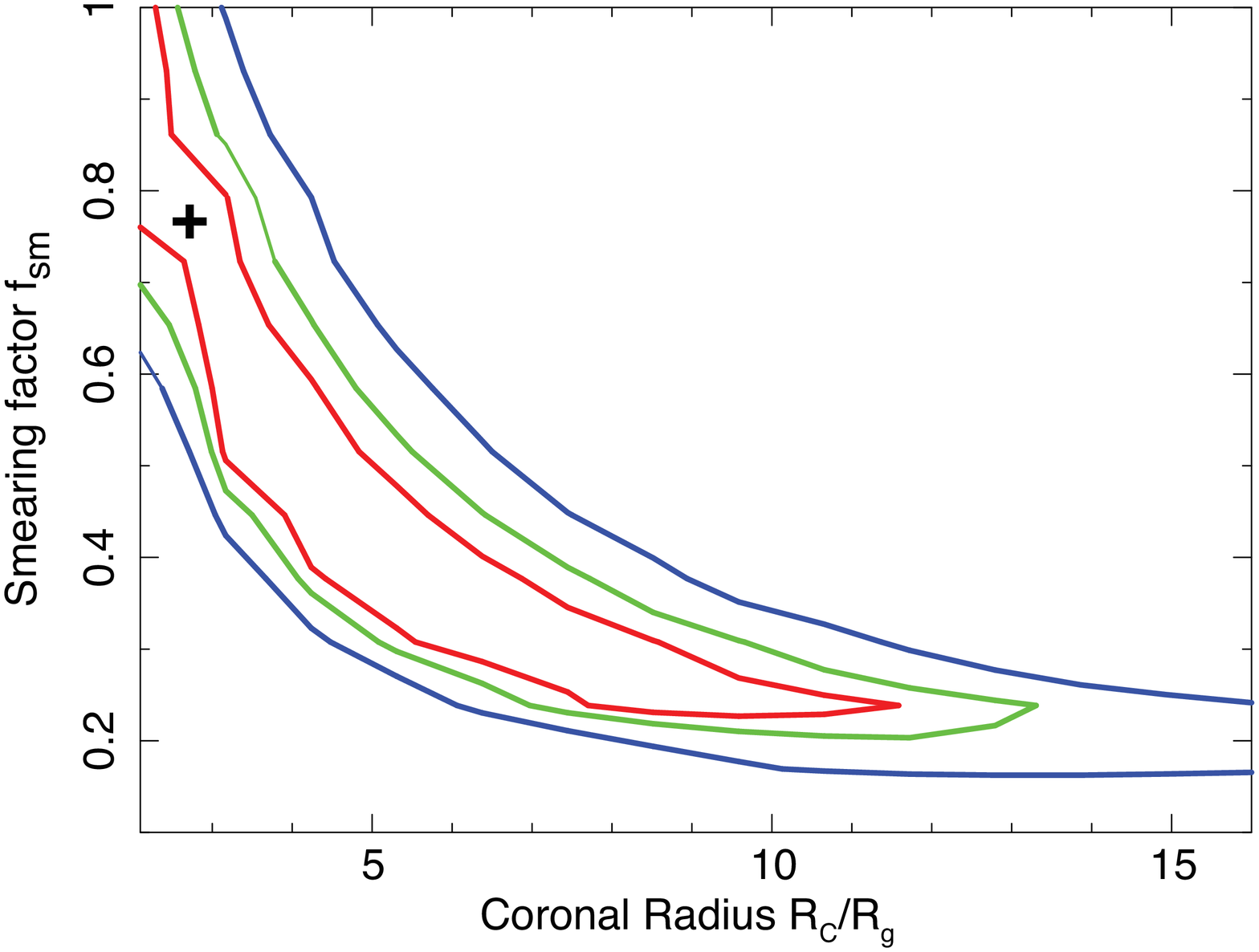}
\end{array}$
\end{center}
\caption{(a) A composite {\it XMM-Newton}/EPIC-pn/mos (black) and {\it NuSTAR} (red and green) spectrum of PDS~456 for different model parameters $(R_c/R_g,f_{\rm sm})$; $(10, 0.76)$ in red, bestfit model in dark, $(2.63, 1)$ in brown, $(2.63, 0.2)$ in blue and no wind case in green. (b) Contour map for the coronal radius $R_c$ and smearing factor $f_{\rm sm}$ (confidence levels at 68\% in red, 90\% in green and 99\% in blue) clearly favoring a compact coronal region with $R_c/R_g \lesssim 10$. } \label{fig:wind2}
\end{figure}

\subsection{Spectral Fit to Fe K UFOs in PDS~456 }

Following the same analysis step discussed in F18, we proceed to construct a grid of model spectra
with a focus on \fexxvi\ in this work. 
Based on the results obtained in F18, we adopt the earlier derived bestfit value of 
$R_T \simeq 8$.
With $\chi^2$ statistics, we find that the bestfit spectrum model is described by a single power-law of photon index $\Gamma=2.12_{-0.008}^{+0.01}$ supplemented independently by the P-Cygni feature (see N15 and F18) in the form of a Gaussian emission at energy of $E_{\rm P-Cygni} = 6.43_{-0.23}^{+0.22}$ keV, and an absorption component is well reproduced by our MHD model. We have adopted  the galactic absorption using {\tt tbabs} where $N_H^{\rm Gal} = 2.4
\times 10^{21}$ cm$^{-2}$ \citep[][]{Kalberla05}. 
%
The similarly modeled Ly$\beta$ line exhibits insignificant signature due to its lower oscillator strength as expected.

%
%
We obtained the bestfit model of $n_{10}=17.3_{-6.1}^{+2.7}, R_c/R_g=2.63_{-0.53}^{+10.9}$ and $f_{\rm sm}=0.76_{-0.48}^{+0.24}$ yielding  $\chi^2$/dof of $369.69/429$ (see {\bf Table~1}). In {\bf Figure~3a} we show the bestfit spectrum in comparison with  data for various parameters. 
It is inferred that small coronal size (i.e. $R_c/R_g \lesssim 10$) is clearly favored by data, while the smearing factor $f_{\rm sm}$ must accordingly vary. The larger the coronal size, the larger the  toroidal motion contribution tends to be. For this reason, the smearing factor needs to be more significant to suppress unnecessarily large line broadening when the corona is relatively large. Lastly, we have alternatively relaxed the parameter values for both the continuum and the P-Cygni emission line components, and the end result is found to be only weakly sensitive to those parameters.  Our spectral modeling thus strongly favors a compact coronal region responsible for photoionization if the line broadening is indeed associated with the wind's transverse  motion as considered here. 

%


\begin{deluxetable}{l|cc}
\tabletypesize{\small} \tablecaption{Model Parameters of the MHD-Wind} \tablewidth{0pt}
\tablehead{Primary Parameter & Range to be Considered & Bestfit Value}
\startdata
Power-Law Photon Index $\Gamma$  & - & $2.12_{-0.008}^{+0.01}$ $^a$  \\
$E_{\rm P-Cygni}$ [keV] & - & $6.43_{-0.24}^{+0.22}$ $^a$ \\ \hline
Wind density$^b$ at the foot point $n_{10}$  & $0.01 - 40$ & $17.3_{-6.1}^{+2.7}$  \\
Radius of Corona $R_c/R_g$ & $2 - 16$ & $2.63_{-0.53}^{+10.9}$ \\
Smearing factor $f_{\rm sm}$ & 0.1 - 1.0  & $0.76_{-0.48}^{+0.24}$ \\ \hline
$\chi^2$/dof & - & 369.69/429 \\
\enddata
\vspace{-0.0in}
\label{tab:tab1}
\begin{flushleft}
We assume $M = 10^9 \Msun$ \citep{Reeves09}, $\theta_{\rm obs}=50\deg$ and $p=1.2$ (from F18).
\\
$^a$ Fixed. \\
$^b$ Wind density normalization at the launching site in units of $10^{10}$ cm$^{-3}$.
\\
\end{flushleft}
\end{deluxetable}

\subsection{Constraint on Coronal Size and Wind's Transverse Motion}

As  illustrated in {\bf Figure~1}, the larger the coronal region, the more the line broadening in the spectrum due to the transverse motion. The line, however, would be too broad when the corona is too large, as implied in this work. For the gradient of $v_\phi(r,\theta)$ over a finite angle subtended by the illuminated wind at radius $r$,  we have parameterized the effective velocity width $\Delta v_{\rm eff}$ by introducing the smearing factor $f_{\rm sm}$ satisfying $0.1 \lesssim f_{\rm sm} \lesssim 1$ as described in \S 2.2. In the context of the present UFO model, a large corona that is radially extended beyond $R_c/R_g \gtrsim 10$ is not consistent with the observed UFO width (see {\bf Figure~3b}) regardless of the smearing factor. We have found in F18 that the disk is favored to be truncated  at $r/R_g \simeq 16$ to achieve the bestfit spectrum for the same data, which is indeed consistent with the present modeling implying a compact corona of $R_c/R_g \lesssim 10$.

\section{Summary \& Discussion}

We have revisited the archetypal Fe K UFOs observed in PDS~456 in an attempt to account for the broad line width that is usually attributed to internal turbulence in the wind. We adopted the same template spectral models in the MHD accretion disk wind scenario used in F18 by further exploiting the generic feature of the MHD-driven disk-winds; i.e. substantial transverse motion around a central corona. In this view, we incorporated toroidal rotation of the wind over a finite angle subtended from a compact corona at the location of the  plasma. Therefore, the size of the corona determines the modulation of the toroidal velocity of the wind that is under direct coronal illumination (see the blue region in {\bf Fig.~1}). The velocity dispersion in turn contributes to broaden the absorption line. In this paper, we thus link the line width to the wind's rotation which is related to the coronal size, thus the UFO line width is used as a proxy to measure the size of the compact corona around the central BH. 
Note that the coronal geometry should affects little in photoionization and kinematics of the UFOs since it is practically point-like compared to the distance where \fexxv/\fexxvi\ are formed.

We have demonstrated by spectral modeling of the 2013/2014 {\it XMM-Newton/NuSTAR} data that the observed \fexxvi\ UFO line width of \obj\ can be potentially used as a measure to infer the central coronal size using the transverse motion within the framework the magnetically-driven disk-winds. Following our earlier spectral modeling in F18, the present  model strongly implies a compact corona of radius of $R_c/R_g \lesssim 10$ with the transverse velocity shift in the wind $\Delta v_{\rm eff}(r) \simeq 0.76 \Delta v_{\rm rms}(r)$ (see {\bf Table~2} and {\bf Figure~3}). As a  unique feature of the MHD winds, the toroidal motion can be substantial especially at smaller radii around the BH as a distinct property compared with the other driving mechanisms. 
Our modeling suggests that a corona of only a few gravitational radii is large enough to broaden the line up to what has been observed.
%
The current spectral modeling, an extension from F18, has  indicated that the wind's rotational motion can indeed account for the broad line width.  

It has been reported by observations of micro-lensed QSOs that X-ray-emitting (coronal) region in QSOs is indeed compact \citep[$r/R_g \lsim 10$; e.g.][]{Morgan08,MacLeod15,Chartas17} consistent with our result.
More specifically, the observed broad iron lines of Seyfert 1s  seem to require a very steep emissivity (e.g. $\epsilon \propto r^{-6}$) perhaps indicating a compact hard X-ray source as in coronae \citep[e.g.][]{FK07,Wilkins12,Dauser13}, while the observed disk reflection component also often  requires a small-scale ``lamp-post" corona \citep[e.g.][]{Garcia14}. 
%
These  observations and modeling  hence necessarily point independently to compactness of a coronal region near the central BH. 
The property of AGN coronae has been studied explicitly with {\it NuSTAR} data by constraining the high-energy cut-off in the hard X-ray continuum, independently suggesting a small coronal size (e.g. $r/R_g \sim 3-10$) for a number of AGNs \citep[e.g.][]{Fabian15,Kamraj18}.

Despite intensive theoretical efforts in the past years to better understand the fundamental role of a global magnetic field in the context of black hole accretion and outflows, there has been a lack of information from spectral modeling in order to securely disentangle various wind driving mechanisms. 
More detailed spectral structures can be revealed by the prospective future X-ray missions, such as {\it XRISM} and {\it Athena}, which will further help probe the width and ionization structure of the outflows, possibly exploiting absorption line spectroscopy as a technique to map the innermost regions in AGNs. 


\acknowledgments 
%
We are grateful to the anonymous referee for the constructive comments. This work is supported in part by 
NASA/ADAP (NNH15ZDA001N-ADAP) and 
{\it Chandra} AO-20 archival proposal grants.

\end{document}